\documentstyle[aps,epsfig]{revtex}

\newcommand{\be}{\begin{equation}}
\newcommand{\ee}{\end{equation}}
\newcommand{\bea}{\begin{eqnarray}}
\newcommand{\eea}{\end{eqnarray}}
\newcommand{\eps}	{\epsilon}
\newcommand{\epe}       {\eps'/\eps}

\begin{document}

\draft
\preprint{
\begin{tabular}{l}
\hbox to\hsize{\hfill KIAS-P00059}\\[-2mm]
\hbox to\hsize{           \hfill SNUTH 00-016}\\[-3mm]
\end{tabular} 
}

\title{
Enhancement of $\eps'/\eps$ 
in the SU(2)$_L \times$SU(2)$_R \times$U(1) model
}

\author{
Ji-ho Jang$^{1,2}$, 
Kang Young Lee$^{3}$,
\thanks{kylee@kias.re.kr},
Seong Chan Park$^4$,
and H.S. Song$^{1,4}$
}

\address{
$^{1}$ Center for Theoretical Physics, 
     Seoul National University, Seoul 151-742, Korea \\
$^{2}$ Institute of Photonics, Electronics, and Information Technology,\\
Chonbuk National University, Chonju Chonbuk 561-756, Korea\\
$^{3}$ School of Physics,
Korea Institute for Advanced Study, Seoul 130-012, Korea\\
$^{4}$ Department of Physics,
Seoul National University, Seoul 151-742, Korea
}

\date{\today}
\maketitle

\begin{abstract}

We explore the possible enhancement of direct CP violating
parameter $\epe$ in the general left-right model
based on the SU(2)$_L \times$SU(2)$_R \times$U(1) gauge group.
The mixing matrix of right-handed quarks, 
$V_{CKM}^R$ is observable in the left-right model, 
and provides new source of CP violating phase.
We calculate the parameter $\epe$ in the left-right model
and show that the new phases from $V_{CKM}^R$ can yield 
the sizable contribution to the direct CP violation
enough to satisfy the recent measurements of Re$(\epe)$
from Fermilab KTeV and CERN NA48 experiments. 

\end{abstract}



\tightenlines

\section{Introduction}

The quantity Re ($\epe$) is a measure of direct CP violation 
in the neutral kaon system.
The experimental situation of Re $(\epe)$
becomes settled down by the recent measurements 
by KTeV \cite{ktev} and NA48 \cite{na48} collaborations.
The present world average \cite{na48} including the earlier NA31 \cite{na31} 
and E731 \cite{e731} results reads
\be
{\rm Re} \left( \epe \right) = (19.3 \pm 2.4) \times 10^{-4},
\ee
which leads to the conclusion that the parameter Re($\epe$) 
is non-zero and
ruled out is the superweak model involving no direct CP violation.
More accurate value of $\epe$ will be obtained 
as NA48 and KTeV experiments further proceed and a new experiment, KLOE,
at the Frascati $\Phi$ factory has started \cite{kloe}.

The theoretical prediction of the standard model (SM) originated in 
the Cabbibo-Kobayashi-Maskawa (CKM) phase is still controversial.
Recently Pallante and Pich have pointed out that the 
final state interactions make it possible for the SM prediction 
to be fitted with the currently measured values of 
Re $(\epe)$ \cite{pallante},
while the earlier predictions shows
more than 2-$\sigma$ deviation from the present measurements \cite{sm}.
Since the hadronic matrix elements have large theoretical
uncertainties, however, it is not settled down yet
that the measured $\epe$ is originated only by the SM.
Moreover, it is well known that the baryogenesis of our universe
need a CP violation beyond that given by the SM.
Therefore it is interesting to consider the new source of
CP violation from new physics beyond the SM
and its implication on $\epe$.
Supersymmetric contribution to Re($\epe$) have been extensively studied
in literatures \cite{susy}
and other models are also attempted \cite{he,wu}.

The left-right (LR) model based on SU(2)$_L \times$ SU(2)$_R \times$ U(1)
gauge group is one of the natural extensions of the SM \cite{LR}.
In the LR model, the right-handed CKM matrix $V^R_{CKM}$ 
which describes mixing of right-handed quarks
is an observable quantity while it is not observable in the SM.
If we make a demand of manifest symmetry 
between left- and right-handed sectors,
$V^R_{CKM}$ should be identical to the usual CKM matrix. 
Then effects of the right-handed current interaction are suppressed 
by a large mass of heavier charged gauge boson $W_R$
and we cannot expect a sizable contribution 
to the CP violating phenomena from the right-handed sector.
Assigning no left-right symmetry manifested,
meanwhile, $V^R_{CKM}$ contains 3 mixing angles and 6 phases in general,
and it may result in exotic CP violations in various processes.
The kaon decay amplitudes generically accompany
the product of CKM matrix elements $\lambda_i = {V_{is}}^* V_{id}$ 
with $i=u,c,t$.
In the SM, the CP phase dominantly appears 
in (13) elements of the CKM matrix for the kaon system 
and the CP violating effects are suppressed by 
smallness of $|V_{td}| \sim 10^{-3}$
in spite of the order 1 phase, $\delta_{CKM}$.
Thus it is possible to enhance the CP violating effect
in the general LR model
if $|\lambda_t^R| \gg |\lambda_t^L|$, 
although suppressed by $m_{W_R}$.

In this work we consider the general version of the LR model
assuming no symmetry between left- and right-handed sectors.
We find that the enhancement of Re($\epe$) in the LR model is 
consistent with the present experiments and 
show the corresponding parameter space. 
This paper is organized as follows:
In section 2, the basic formalism of
the $\Delta S =1$ effective Hamiltonian in the LR model
is presented.
We calculate the contribution of the right handed sector to
the parameter $\epe$ in section 3
and perform the numerical study
under the constraints from measured $\eps_K$ and $\Delta m_K$
in section 4.
Finally we conclude in section 5.

\section{$\Delta S = 1$ effective Hamiltonian in the LR model}

The $K \to \pi \pi$ processes are described by
the $\Delta S =1$ effective Hamiltonian written by
\bea
{\cal H}_{eff}^{\Delta S =1} &=& \frac{G_F}{\sqrt{2}} 
    \left[ ~\sum^{2}_{i=1} \left( 
           \lambda_u^{_{L}} C_i(\mu) Q_i(\mu) 
         + \lambda^R_u C'_i(\mu) Q'_i(\mu) \right)
      \right.
\nonumber \\
   && \left. ~ -\sum^{10}_{j=3} \left( 
           \lambda_t^{_{L}} C_j(\mu) Q_j(\mu) 
         + \lambda^R_t C'_j(\mu) Q'_j(\mu) \right) \right]
\eea
where $Q_i$ are the SM operators,
$Q'_i$ their chiral conjugate operators,
and $C_i$ and $C'_i$ are corresponding Wilson coefficients.
We follow the convention of Ref. \cite{buras0,buras1}
for the explicit form of operators $Q_i$.
For simplicity, we set the left-right mixing to be zero in this work.
The new Wilson coefficients $C'_i$ at the scale $\mu = m_{W_R}$
are determined by matching the Feynman diagrams with $W_R$ boson exchanges 
to the effective Hamiltonian (2).
The relevant Feynman rules involved in the diagrams of
$\gamma$-penguin and gluon-penguin are obtained by
the (L $\leftrightarrow$ R) exchange of their chiral conjugates of the SM.
Consequently we have the effective $\gamma$- and gluon-penguin
vertices as
\bea
( \bar{s} \gamma d )_R &=& -i \lambda_i^R \frac{G_F}{\sqrt{2}} \frac{e}{8\pi^2}
            \frac{g_R^2}{g_L^2} D_0(x'_i) 
          \bar{s} (q^2 \gamma_\mu - q_\mu \not{q}) (1+\gamma_5) d,
\\
(\bar{s} G^a d )_R &=& -i \lambda_i^R \frac{G_F}{\sqrt{2}} \frac{g_s}{8\pi^2}
            \frac{g_R^2}{g_L^2} E_0(x'_i) 
          \bar{s}_\alpha (q^2 \gamma_\mu - q_\mu \not{q}) 
                          (1+\gamma_5) T^a_{\alpha \beta} d_{\beta},
\eea
where ${x_i}' = m_{i}^2/m_{W_R}^2$ and $i=u,c,t$. 
The loop functions $D_0(x)$ and $E_0(x)$ are
same as those of the SM and given in the Ref. \cite{buras1}.
Besides 
the effective $Z$-penguin vertex 
with the internal $W_R$ boson exchanges is given by
\be
(\bar{s} Z d)_R = i \lambda_i^R \frac{G_F}{\sqrt{2}} \frac{e}{2\pi^2}
            m_{_Z}^2 
            \frac{\cos \theta_{W}}{\sin \theta_{W}} 
            \frac{g_R^2}{g_L^2} 
             C_R(x'_i) 
            \bar{s} \gamma_\mu (1+\gamma_5) d,
\ee
with the new loop function  
\be
C_R(x) = \frac{x}{16} \left( \frac{3}{1-x}
               + \frac{4-2x+x^2}{(1-x)^2} \ln x \right),
\ee
in the leading order of the suppression by $m_{W_R}$.
Since the $Z f \bar{f}$ and $Z W^+ W^-$ vertex is left-right asymmetric,
the loop function of $(\bar{s} Z d)_R$ vertex 
is different from that of the $Z$-penguin vertex in the SM.
We do not include the contributions from the charged Higgs boson here,
which is acceptable because we have a freedom to let the charged Higgs boson 
be sufficiently heavy.

The relative strength of new effective vertices to
the SM ones are generically
given by
\be
\left| \frac{(\bar{s} V d)_{R}}{(\bar{s} V d)_{SM}} \right|
           = \left| \frac{\lambda_i^R}{\lambda_i^L} \right|  
             \frac{g_R^2}{g_L^2} 
             \frac{F_V^R(x'_i)}{F_V^L(x_i)},
\ee
where $F(x)$ is a generic loop function.
Actually the generic suppression factor involved 
in the right-handed sector is given by
$\beta_g \equiv (g_R^2/g_L^2)(m^2_{{W_{L}}}/m^2_{{W_{R}}})$ in the LR model. 
Note that the ratio of Eq. (7) goes to the generic suppression factor
$ \sim |\lambda_i^R/\lambda_i^L| \beta_g$,
if $F(x) \sim x$.
We point out that the new vertices can be enhanced 
by the factor $| \lambda_i^R/\lambda_i^L |$ 
while suppressed by the large mass of $W_R$.
The loop functions $D_0(x')$, $E_0(x')$, and $C_R(x')$ behave
like logarithmic functions when $x' \ll 1$ so that 
the suppression by the $m_{W_R}$ is weaker than $\beta_g$.

We also have a extra neutral gauge boson in the LR model 
and the corresponding $\bar{s} Z' d$ vertex.
Due to the $Z'$ propagator, its contribution 
to the Wilson coefficients is additionally suppressed 
by the factor of $m^2_{Z}/m^2_{{Z'}} < 0.02$ 
from the present bound on $m_{Z'}$ \cite{zprime}.
Hence we ignore $\bar{s} Z' d$ vertex in this work.

There are 2 kinds of box diagram
contributing to the terms $C'_i Q'_i$;
one has two $W_R$ exchange and the other one $W_R$ and 
one ordinary $W_L$ exchanges. 
The box diagram containing two $W_R$ bosons is computed as
\be
\left[ (\bar{s} d) (\bar{q} q) \right]_{RR}
= -i \lambda^R_i \frac{G_F}{\sqrt{2}} \frac{g_L^2}{4\pi^2}
      | V_{jq}^R |^2 \beta_g \frac{g_R^2}{g_L^2} \tilde{B}({x'}_i,{x'}_j)
                           (\bar{d} P_L \gamma_\mu s ) 
                           (\bar{q} P_L \gamma^\mu q ), 
\ee
where the function $\tilde{B}({x'}_i,{x'}_j)$ ($i,j = u,c,t$)
is found in Ref. \cite{buras1}.
We find that this contribution is additionally suppressed by $\beta_g$
as well as the loop function factor 
$(g_R^2/g_L^2) (\tilde{B}({x'}_i,{x'}_j)/\tilde{B}(x_i,x_j) )$.
For the box diagram with one $W_R$ boson and 
one $W_L$ boson exchanges,
the chiral structure makes the contributions 
proportional to the masses of internal quarks
as well as the CKM factors such that
$\left[ (\bar{s} d) (\bar{d} d) \right]_{LR} \propto 
\lambda^R_i | V_{id}^L |^2 m_i^2$.
Thus this is additionally suppressed by
the mass ratio $m_i^2/m_W^2 \sim (10^{-4} - 10^{-8})$ ($i=u,c$) 
or the CKM factor $ | V_{id}^L |^2 \sim 10^{-5}$ ($i=t$).
Consequently the contributions from box diagrams are much smaller
than those from penguin diagrams and not considered in this work.

Matching the full theory to the effective theory given by the Eq. (2)
at the scale $\mu = m_W$,
we have the Wilson coefficients at next-to-leading order (NLO)
\bea
C'_1(m_W) &=& \frac{11}{2} \frac{\alpha_s(m_W)}{4 \pi} \beta_g
\nonumber \\
C'_2(m_W) &=& \left( 1 - \frac{11}{6} \frac{\alpha_s(m_W)}{4 \pi}
                - \frac{35}{18} \frac{\alpha}{4 \pi} \right) \beta_g
\nonumber \\
C'_3(m_W) &=& - \frac{\alpha_s(m_W)}{24 \pi} 
                \frac{g_R^2}{g_L^2} \tilde{E}_0(x'_t) 
\nonumber \\
C'_4(m_W) &=&  C'_6(m_W) 
           = \frac{\alpha_s(m_W)}{8 \pi} 
                \frac{g_R^2}{g_L^2} \tilde{E}_0(x'_t)
\nonumber \\
C'_5(m_W) &=& \left( - \frac{\alpha_s(m_W)}{24 \pi} \tilde{E}_0(x'_t)
              + \frac{1}{\sin^2 \theta_W} \frac{\alpha}{6 \pi} C_R(x'_t)
              \right) \frac{g_R^2}{g_L^2}
\nonumber \\
C'_7(m_W) &=& \frac{\alpha}{6 \pi} \frac{g_R^2}{g_L^2} 
              \left[ 4 C_R(x'_t) + \tilde{D}_0(x'_t) 
                       - \frac{4}{\sin^2 \theta_W} C_R(x'_t) \right]
\nonumber \\
C'_9(m_W) &=& \frac{\alpha}{6 \pi} \frac{g_R^2}{g_L^2} 
              \left[ 4 C_R(x'_t) + \tilde{D}_0(x'_t) \right]
\nonumber \\
C'_8(m_W) &=& C'_{10}(m_W) = 0.
\eea
We ignore the running from the scale $\mu = m_{W_R}$ to $ m_{W}$ 
for simplicity and perform the matching only at $\mu = m_{W}$. 
The complete renormalization group (RG) evolution 
of the Wilson coefficients 
$(C_i,C'_i)$ from the scale $\mu = m_{W}$ to $\mu = m_c$ is 
governed by a $20 \times 20$ anomalous dimension matrix.
Since the strong interaction preserves chirality,
new operators $Q'_i$ are not mixed with the SM operators
and evolved separately.
Thus the anomalous dimension matrix is decomposed into two
$10 \times 10$ matrices which are identical to each other.
The $10 \times 10$ anomalous dimension matrix has been calculated 
by several authors at NLO \cite{adm}.
Here we use the numerical values listed in Ref. \cite{buras0},
obtained under NDR scheme.
Finally the values of the Wilson coefficients 
at the scale $\mu = m_c$ are determined
after solving the RG equation.

\section{$\epe$ in the LR model}

The complex parameter $\eps'$ is defined as
\be
\eps' = \frac{1}{\sqrt{2}} 
        {\rm Im} \left( \frac{A_2}{A_0} \right) 
          e^{i (\pi/2 + \delta_2-\delta_0)} ,
\ee
where $A_{I=0,2}$ are the isospin amplitudes in $K \to \pi \pi$
decays
and $\delta_{0,2}$ are the corresponding strong phases.
The ratio Re $(\epe)$ 
is obtained from the measured ratio 
$\eta_{00} \equiv A( K_L \to \pi^0 \pi^0)/A( K_S \to \pi^0 \pi^0)$
and
$\eta_{\pm} \equiv A( K_L \to \pi^+ \pi^-)/A( K_S \to \pi^+ \pi^-)$
as
\be
\left| \frac{\eta_{00}}{\eta_{\pm}} \right|^2 \approx
   1 - 6~\mbox{Re} \left( \frac{\epsilon'}{\epsilon} \right) ,
\ee
where the deviation of $| \eta_{00}/\eta_{\pm} |$ from 1
indicates the direct CP violation in $K \to \pi \pi$ decays.
Using the Hamiltonian in Eq. (2),
we can express the parameter Re $(\epe)$ with the CKM factors;
\be
{\rm Re} \left( \frac{\epsilon'}{\epsilon} \right) 
          =  N_t^L ~{\rm Im}~ \lambda_t^L
           + N_t^R ~{\rm Im}~ \lambda_t^R
           - N_u^R ~{\rm Im}~ \lambda_u^R, 
\ee
where the coefficients $N_i^{L(R)}$s are given by
\be
N_i^{L(R)} = \frac{G_F \omega}{2 |\epsilon| \rm{Re}A_0}
             \left( \sum_i C_i^{(\prime)} \langle Q_i^{(\prime)} \rangle_0
             -\frac{1}{\omega}
               \sum_i C_i^{(\prime)} \langle Q_i^{(\prime)} \rangle_2 \right),
\ee
in terms of the evolved Wilson coefficients $C_i^{(\prime)}(\mu=m_c)$ and
hadronic matrix elements $\langle Q_i^{(\prime)} \rangle_{0,2}$.
The explicit form of hadronic matrix elements 
$\langle Q_i \rangle_I$
are listed in Ref. \cite{buras1,buras2}.
The parameter $\omega$ is defined by the ratio of isospin amplitudes,
as $\omega \equiv {\rm Re}~A_2/{\rm Re}~A_0$.
The ($-$) sign of the last term in Eq. (12) is owing to
the convention of the effective Hamiltonian of Eq. (2).
The value of $N_t^L$ predicted in the SM is reduced 
by cancellation between 
$\Delta I = 1/2$ and $\Delta I = 3/2$ contributions,
as the electroweak penguin contributions are enhanced
by large top quark mass.
In the LR model, $W_R$ is much heavier than the top quark
and the electroweak penguin contributions are relatively small
as $x'_t \ll 1$. 
Thus there is less cancellation between 
$\Delta I = 1/2$ and $\Delta I = 3/2$ contributions in this model,
which yields enhancement of the coefficients $N_i^R$.
We notify that $\lambda_u^R$ may be the complex number, 
and contributions of the right-handed sector
to Re $(\epe)$ depend on two CKM parameters
Im $\lambda_t^R$ and Im $\lambda_u^R$ as a result, 
while the SM contribution consists of the Im $\lambda_t^L$ alone.
It leads that a direct bounds on right-handed CKM elements
are hardly obtained.

\section{Numerical Study}

It is well-known that the $K-\bar{K}$ mixing puts 
stringent constraints on the LR model \cite{beall}.
The parameter $\eps_K$ and the mass difference $\Delta m_K$
is obtained from the off-diagonal element $M_{12}$ in the neutral kaon
mass matrix.
The leading contribution of the LR model to $M_{12}$
comes from the box diagram with one $W_L$ and one $W_R$ gauge bosons
as internal lines.
One can find the relevant $\Delta S =2$ effective Hamiltonian
and related formulae in the Ref. \cite{nishiura}.
If the left-right symmetry manifests, 
the mass of $W_R$ boson should be greater than 1.6 TeV \cite{barenboim1} 
to satisfy the experimental $\Delta m_K$ and $\epsilon_K$ data
\cite{buras1,pdg},
\bea
&&\Delta m_K = 3.51 \times 10^{-15}~{\rm GeV},
\nonumber \\
&&\epsilon_K = (2.280 \pm 0.013) \times 10^{-3} . 
\eea
The bound on $m_{W_R}$ can be lowered
by assuming $V^R_{CKM} \ne V^L_{CKM}$ 
\cite{london,kurimoto} and it is the case considered here.

In the general LR model, the gauge couplings of SU(2)$_L$
and SU(2)$_R$ groups are not necessarily same 
but expected to be of the same order 
to avoid the fine-tuning and maintain the perturbativity.
For the numerical analysis, we let $g_L^2/g_R^2 =1$ here.
We limit the contribution of the LR model
to the $\epsilon_K$ to be within the measured error, 
which implies that the indirect CP violation is originated 
principally by the left-handed sector.
The box diagrams of the SM is known to describe about 70\% of 
the measured $K_L - K_S$ mass difference  
and the remaining part is attributed to unknown contributions
including nonperturbative effects.
So we assume that the new contributions $\Delta m_K^R$ is required not to
exceed 30\% of the measured $\Delta m_K$ here.

The vacuum insertion method with $B_i^{(\Delta I)} = 1$
provides a good approximation of $\epe$ in the SM.
But the parameters $B^{(1/2)}_1$, $B^{(1/2)}_2$, $B^{(3/2)}_1$
show large deviations from the values expected in
the vacuum saturation limit.
In the SM, $B_{i=1,2}$ do not play roles in CP violation
since Im $\lambda_u^L$ is extremely small.
However, the term Im $\lambda_u^R$ from the right-handed sector
is not necessarily small
and it can contribute to $\epe$ considerably.
Here, we consider the special case of ${\rm Im}~\lambda_u^R=0$
for the time being
in order to investigate the enhancement of $\epe$.
We calculate the ratio
$R \equiv \mbox{Re}(\epe)_{R} /\mbox{Re}(\epe)_{SM}$
in the vacuum saturation limit 
and plot it with varying $m_{W_R}$ from 800 GeV to 2 TeV in Fig. 1.
Constraints by the $\Delta m_K$ and $\epsilon_K$ data are considered.
This plot indicates the enhancement by the CKM factor
${\rm Im}~\lambda_t^R /{\rm Im}~\lambda_t^L$
since contributions to Re($\epe$) come from the CKM elements
${\rm Im}~\lambda_t^{L,R}$ alone in this case.
We show the dependence of $\epe$ on ${\rm Im}~\lambda_t^{R}$
with respect to $ m_{W_{_R}}$ in Fig. 2.

For the realistic prediction without the assumption that
${\rm Im}~\lambda_u^R=0$, 
we calculate $\epe$ by scanning all the angles and phases of $V^R_{CKM}$
which are constrained by the $\Delta m_K$ and $\epsilon_K$ data.
In this paper,
we adopt the values of $B^{(1/2)}_1$, $B^{(1/2)}_2$, and $B^{(3/2)}_1$ 
extracted in Refs. \cite{buras0,buras1} by a phenomenological approach
\be
B^{(1/2)}_1 = 16.5,~~
B^{(1/2)}_2 = 6.6,~~
B^{(3/2)}_1 = 0.453,
\ee
and those of $B^{(1/2)}_6$ and $B^{(3/2)}_8$
summarized by \cite{buras0,buras1,buras2}
\be
B^{(1/2)}_6 = 1,~~~~
B^{(3/2)}_8 = 1.
\ee
Figure 3 plots the all possible parameter set of 
$( {\rm Im}~\lambda_u^R, {\rm Im}~\lambda_t^R) $
which satisfy the recent $\epe$ data of Eq. (1) at 2-$\sigma$ level 
under the $\Delta m_K$ and $\epsilon_K$ constraints
for $m_{W_R}=800$ GeV, which is close to the present lower bound of 
extra $W$ gauge boson from experiment.
Correlations of the parameters ${\rm Im}~\lambda_u^R$ and
$ {\rm Im}~\lambda_t^R $ with $\epe$ are shown in Fig. 4.
We find that the values of ${\rm Im}~\lambda_t^R$ center on a nonzero value, 
while the values of ${\rm Im}~\lambda_u^R$ are distributed 
centering on zero.
Thus we conclude that Im $\lambda_t^R $ is principally responsible
for the new contributions to the $\epe$. 
We note that there exist a few points far from the accumulated region, 
which indicate the fine-tuned combinations of parameters
and they should be less meaningful.

\section{Conclusion}

We explore the possibility that the LR model provides
sizable direct CP violation of the kaon system
without affecting the $K - \bar{K}$ mixing system.
It implies the scenario that
the $\epsilon_K$ is explained by the SM sector 
and the $\epe$ explained by the right-handed sector.
In conclusion, we show that the recent measurement of $\epe$
is explained in the framework of the general left-right model
without fine-tuning.
We pointed out three possibility that the contributions of the LR model
can be larger than expected; (1) enhancement by CKM factors,
(2) the weaker suppressions for the effective vertices than 
the tree level suppression factor $\beta_g$,
and (3) less cancellation between 
$\Delta I = 1/2$ and $\Delta I = 3/2$ contributions.
On the other hand it is likely for the LR model to give interesting
predictions on other observables, i.e. hyperon CP violation \cite{hyperon},
$K \to \pi \nu \bar{\nu}$ decays etc.
in the parameter space studied here. 
Measurements of these observables enables us to obtain more information
on $V_{CKM}^R$ and test the LR model.

\acknowledgments

We thank P. Ko and S.Y. Choi for their valuable comments.
J.-H.J thank J.E. Kim for hospitality.
This work is supported in part
by the Korean Science and Engineering Foundation (KOSEF) 
through the SRC program of the Center for Theoretical Physics (CTP),
and by the BK21 program of Ministry of Education
at Seoul National University.

\begin{center}
\begin{figure}[htb]
\hbox to\textwidth{\hss\epsfig{file=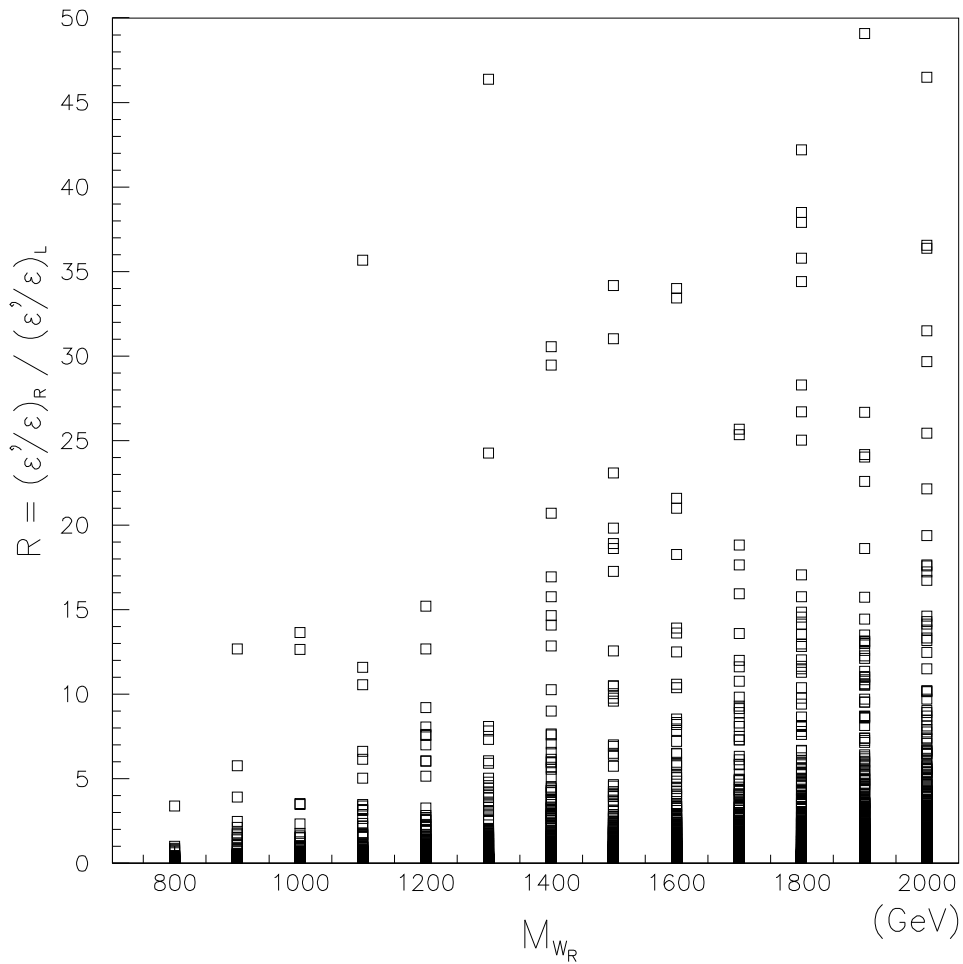,height=15cm}\hss}
\vspace{1cm}
\caption{
The ratio of the $\epe$ from right-handed sector to that of the 
Standard Model with respect to the mass of $W_R$ boson
in the vacuum saturation limit
under the assumption that Im $\lambda_u^R=0$.
}
\end{figure}
\end{center}
 
\begin{center}
\begin{figure}[htb]
\hbox to\textwidth{\hss\epsfig{file=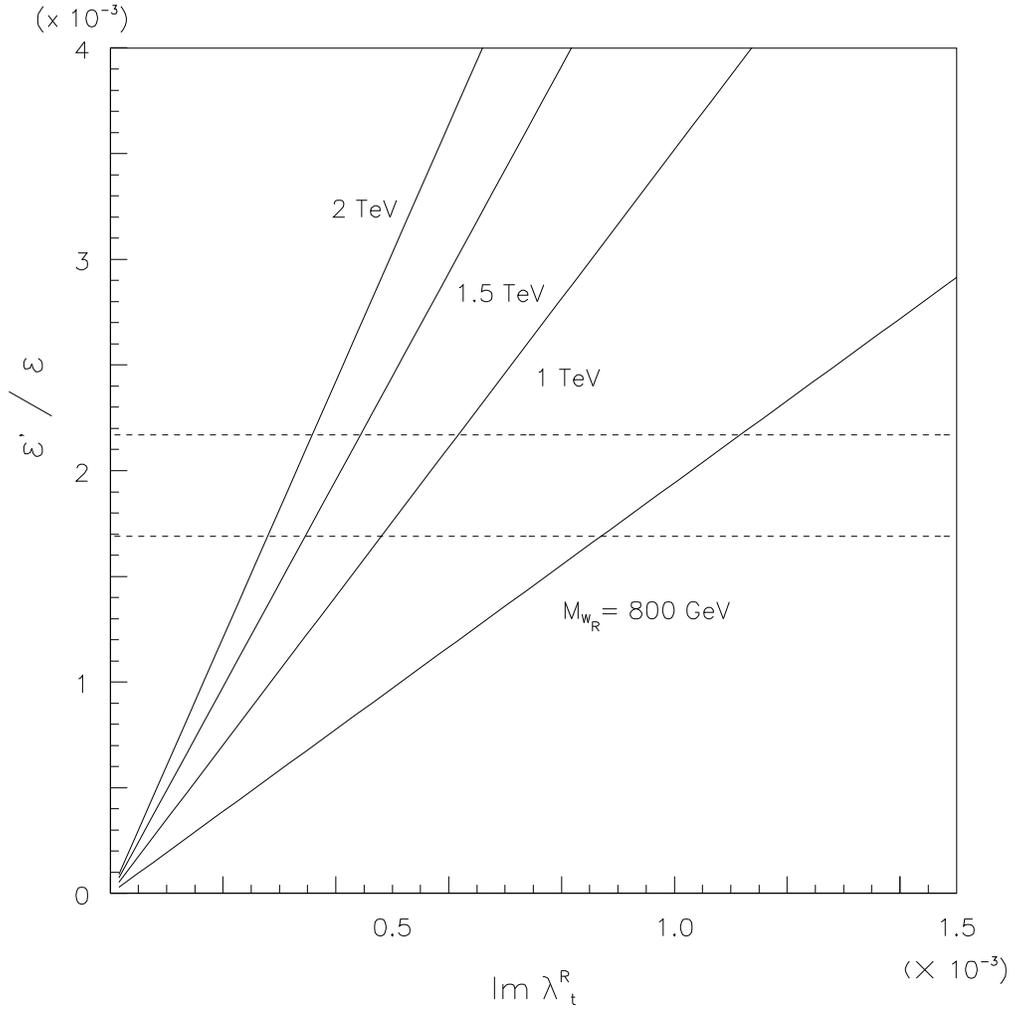,height=15cm}\hss}
\vspace{1cm}
\caption{
Dependences of $\epe$ on the CKM parameter Im $\lambda_t^R$
for each value of $m_{W_{_R}}$
in the vacuum saturation limit
under the assumption that Im $\lambda_u^R=0$.
}
\end{figure}
\end{center}
 
\begin{center}
\begin{figure}[htb]
\hbox to\textwidth{\hss\epsfig{file=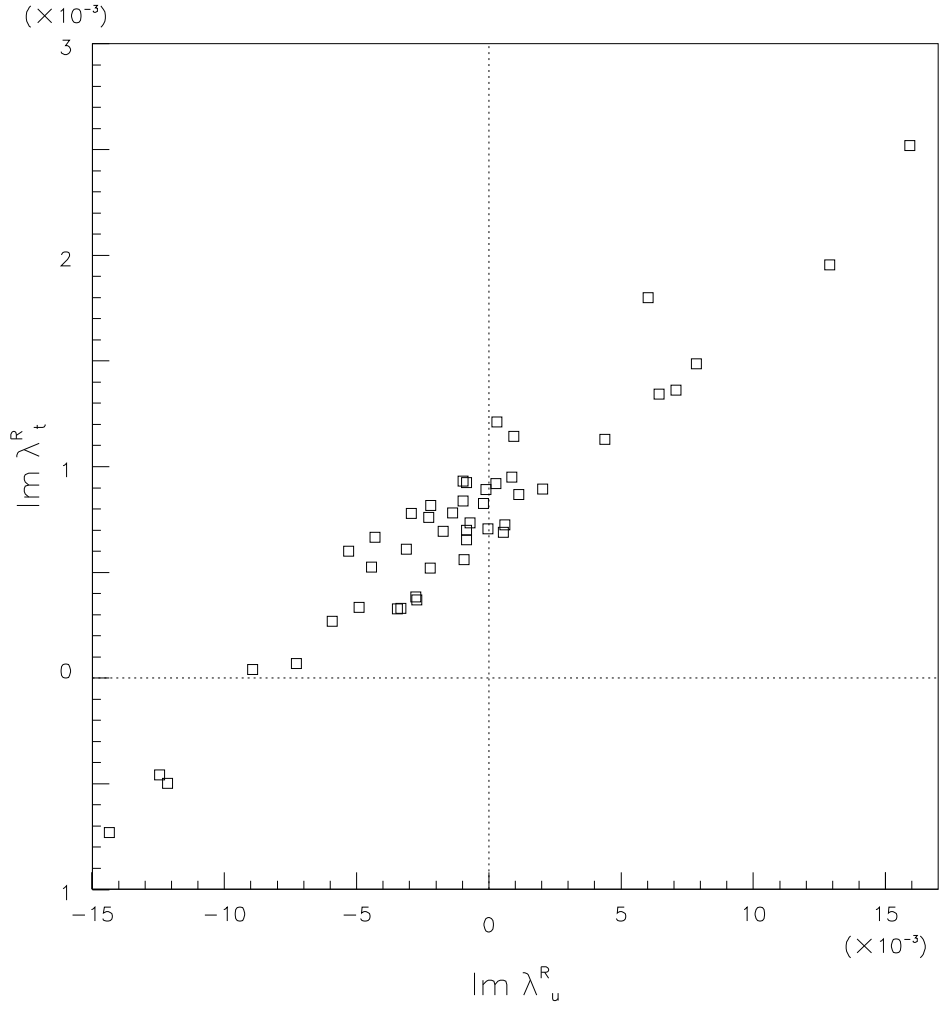,height=15cm}\hss}
\vspace{1cm}
\caption{
Parameter set of $({\rm Im}~\lambda_u^R, {\rm Im}~\lambda_t^R) $
which satisfy the present $\Delta m_K$, $\epsilon_K$  data and
the recent $\epe$ data for  $ m_{W_R} =800$ GeV. 
}
\end{figure}
\end{center}

\newpage
\mbox{ }

\begin{center}
\begin{figure}[htb]
\hbox to\textwidth{\hss\epsfig{file=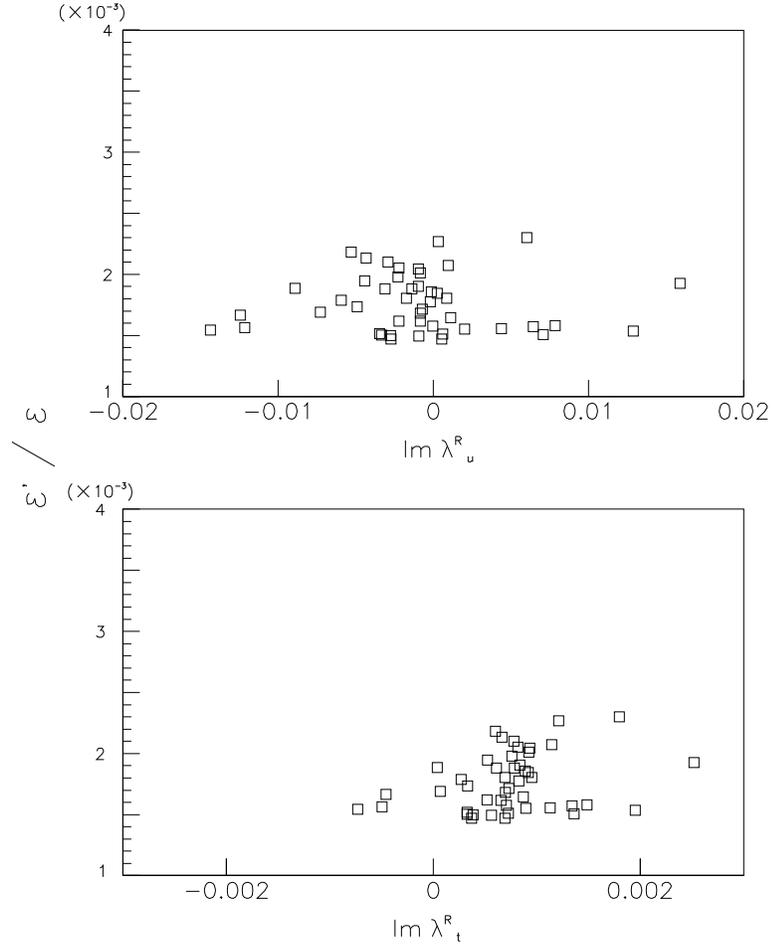,height=15cm}\hss}
\vspace{1cm}
\caption{
Distributions of the $\epe$ prediction with respect to $\lambda_u^R$
and $ \lambda_t^R $ which satisfy the $\Delta m_K$ and $\epsilon_K$ data 
for $ m_{W_R} =800$ GeV.
}
\end{figure}
\end{center}
 
\vfil\eject

\end{document}